\documentclass[11pt]{article}

\usepackage[preprint]{acl}

\usepackage{times}
\usepackage{latexsym}

\usepackage[T1]{fontenc}
\usepackage{amsmath}
\usepackage[utf8]{inputenc}

\usepackage{microtype}

\usepackage{inconsolata}

\usepackage{graphicx}
\usepackage{booktabs}
\usepackage[table]{xcolor}
\usepackage{multirow}
\newcommand{\todo}[1]{\todopurple{TODO: #1}}

\newcommand{\todoc}[2]{{\textcolor{#1}{\textbf{#2}}}}

\newcommand{\todopurple}[1]{\todoc{purple}{\textbf{[#1]}}}

\newcommand{\phenomenon}{\texttt{SLUMP}}
\newcommand{\spec}{\texttt{emergent specification}}

\newcommand{\memory}{\texttt{ProjectGuard}}

\newcommand{\forecaster}{\texttt{Forecaster}}
\newcommand{\good}[1]{\cellcolor{green!12}\textbf{#1}}

\usepackage{listings}
\title{When the Specification Emerges: Benchmarking Faithfulness Loss in Long-Horizon Coding Agents}

\author{Lu Yan \and Xuan Chen \and Xiangyu Zhang \\
Purdue University         }

\begin{document}
\maketitle
\begin{abstract}
Current coding-agent benchmarks usually provide the full task specification upfront. Real research coding often does not: the intended system is progressively disclosed through interaction, requiring the agent to track durable design commitments across a long session. We introduce a benchmark for this setting and study faithfulne\underline{S}s \underline{L}oss \underline{U}nder e\underline{M}ergent s\underline{P}ecification (\phenomenon{}), defined as the reduction in final implementation faithfulness under emergent specification relative to a single-shot specification control. The benchmark contains 20 recent ML papers (10 ICML 2025, 10 NeurIPS 2025), 371 atomic verifiable components, and interaction scripts of approximately 60 coding requests that progressively disclose the target design without revealing the paper itself. Final repositories are scored with a five-level component-faithfulness rubric and accompanied by an exposure audit to verify that scored components are recoverable from the visible interaction. Evaluated on Claude Code and Codex, the single-shot specification control achieves higher overall implementation fidelity on 16/20 and 14/20 papers, respectively. Structural integration degrades under emergent specification on both platforms, while semantic faithfulness loss is substantial on Claude Code and small on Codex. As a mitigation case study, we introduce \memory{}, an external project-state layer for specification tracking. On Claude Code, \memory{} recovers 90\% of the faithfulness gap, increases fully faithful components from 118 to 181, and reduces severe failures from 72 to 49. These results identify specification tracking as a distinct evaluation target for long-horizon coding agents. The benchmark is open-sourced on \href{https://github.com/lunaryan/drift-benchmark}{Github}.
\end{abstract}

\section{Introduction}
\label{sec:introduction}

AI coding agents are increasingly used to implement research ideas in practice.
Recent benchmarks evaluate whether agents can translate papers or task descriptions into working code, including PaperBench~\cite{starace2025paperbench}, MLR-Bench~\cite{chen2025mlrbench}, LMR-Bench~\cite{lmrbench}, and RECODE-H~\cite{miao2025recode}.
These benchmarks have substantially advanced evaluation of coding agents, but they largely share a common assumption: the target specification is available at the start of the session, whether as a paper, a task description, or a test suite.
They do not measure whether an agent can faithfully implement a system whose specification is distributed across a long interaction.

Real research workflows often violate this assumption.
A researcher rarely begins with a complete implementation blueprint.
Early turns frame the idea and compare alternatives, middle turns implement components under partial specification, and later turns refine and integrate those components into a unified training and evaluation pipeline.
The intended method is therefore not written once in a canonical prompt.
It is assembled across many turns, interleaved with exploratory alternatives, local revisions, and integration decisions.
Faithful implementation in this setting requires specification tracking: the agent must retain durable design commitments introduced earlier in the interaction and preserve them as the codebase evolves~\citep{zhu2026swecontextbench,liu2026scalable}.

We study faithfulne\underline{S}s \underline{L}oss \underline{U}nder e\underline{M}ergent s\underline{P}ecification (\phenomenon{}): the reduction in final implementation faithfulness that occurs when the target design is disclosed progressively through interaction rather than provided upfront.
\phenomenon{} is an endpoint-based notion.
We do not attempt to measure temporal drift inside a session.
Instead, we ask whether the final repository produced under emergent specification is less faithful than the final repository produced by the same platform under a single-shot specification control.
We analyze this loss along two diagnostic dimensions: semantic faithfulness to the committed design and structural integration of earlier modules into the final system.

To measure \phenomenon{}, we construct a benchmark from 20 randomly sampled ML papers (10 from ICML 2025 and 10 from NeurIPS 2025).
For each paper, we extract a canonical specification from the original text, decompose it into atomic verifiable components, and derive an interaction script of approximately 60 coding requests that progressively discloses the target design without revealing the paper itself.
Across the 20 papers, this yields 371 frozen components for evaluation.
We score the final repository against a five-level component faithfulness rubric and complement scoring with an exposure audit that checks, for each component, whether it is recoverable from the agent-visible interaction.
This audit is necessary because emergent-specification evaluation is only meaningful if the target components are actually inferable from the visible dialogue.

We instantiate the benchmark on two long-horizon coding platforms, Claude Code and Codex, under two conditions: emergent specification and a single-shot specification control.
The single-shot control achieves higher overall implementation fidelity on 16 of 20 papers on Claude Code and 14 of 20 papers on Codex, providing direct evidence of \phenomenon{}.
The decomposition of this gap differs across platforms: structural integration degrades under emergent specification on both systems, while semantic faithfulness loss is substantial on Claude Code and small on Codex.
These results show that progressive specification disclosure imposes a measurable cost on final implementation fidelity even when the underlying platform and tooling are held fixed.

As a case study in targeted mitigation, we introduce \memory{}, an external project-state layer for specification tracking.
\memory{} maintains a semantic view of committed project knowledge and a structural view of repository organization, then injects a compatibility-aware project brief before each coding turn or proactive restart.
On Claude Code, where \phenomenon{} is substantial, \memory{} recovers 90\% of the faithfulness gap between emergent specification and the single-shot specification control, increases the number of fully faithful components by 53\%, and reduces severe failures by 32\%.

This paper makes three contributions.
First, it introduces a benchmark and evaluation methodology for specification tracking under emergent-specification coding, combining long interaction scripts, component-level faithfulness scoring, and an exposure-based fairness audit.
Second, it provides empirical evidence that emergent specification reduces final implementation fidelity relative to a single-shot specification control on current long-horizon coding platforms.
Third, it presents \memory{} as a targeted mitigation case study, showing that externally maintained project state can substantially reduce \phenomenon{} without modifying the underlying model.

The benchmark is open-sourced on \href{https://github.com/lunaryan/drift-benchmark}{Github}.
\section{Problem Statement}
\label{sec:problem_setting}

\subsection{Emergent-Specification Coding}
\label{subsec:emergent_spec}

We study emergent-specification coding, a multi-turn coding setting in which the target system is not given in a single initial prompt but is progressively disclosed through user requests. Early turns may explore alternatives or defer details, middle turns implement components under partial specification, and later turns refine and integrate those components into a complete system. The implementation target is therefore distributed across the interaction rather than stated upfront.

This setting requires the agent to retain both local task state and project-level specification state. Local task state includes recent files, current errors, and the immediate edit under discussion. Project-level specification state includes durable design commitments such as equations, architectural constraints, interface contracts, and evaluation requirements. Faithful implementation requires tracking both across a long interaction.

We evaluate agents against the final committed design: the set of design facts that are established by the end of the interaction as part of the intended system. Exploratory alternatives, abandoned branches, and unresolved suggestions are not part of the target. The benchmark construction in Section~\ref{sec:benchmark} operationalizes this target using a canonical specification derived from the source paper and an interaction script that reveals it progressively without exposing the paper itself.

\subsection{\phenomenon{}: Faithfulness Loss Under Emergent Specification}
\label{subsec:faithfulness_loss}

Let $p$ denote a task and $a$ an agent platform. Let $R^{\mathrm{em}}_{p,a}$ be the final repository produced when the agent receives the emergent multi-turn interaction, and let $R^{\mathrm{ss}}_{p,a}$ be the final repository produced when the same agent receives the complete specification in a single shot upfront. For any implementation-faithfulness metric $F$, we define faithfulne\underline{S}s \underline{L}oss \underline{U}nder e\underline{M}ergent s\underline{P}ecification (\phenomenon{}) as
\[
\phenomenon{}_F(p,a) = F(R^{\mathrm{ss}}_{p,a}) - F(R^{\mathrm{em}}_{p,a}).
\]
Positive values indicate that progressive disclosure reduces final implementation faithfulness relative to the single-shot specification control.

This definition is deliberately endpoint-based. We do not attempt to measure temporal drift within a session. Instead, we measure whether the final implementation under emergent specification is less faithful than the final implementation obtained when the full design is available from the start.

We analyze \phenomenon{} along two diagnostic dimensions. The first is \textbf{semantic faithfulness}: whether the final code implements the committed algorithmic content of the target design. The second is \textbf{structural integration}: whether later-stage code preserves and reuses earlier modules rather than bypassing or reimplementing them along incompatible paths. These dimensions are measured later using Mean Component Faithfulness (MCF) and Dependency Integration Ratio (DIR), respectively. They are diagnostic views of the observed loss rather than claims about its unique cause.
\begin{figure*}[t]
    \centering
    \includegraphics[width=.77\textwidth]{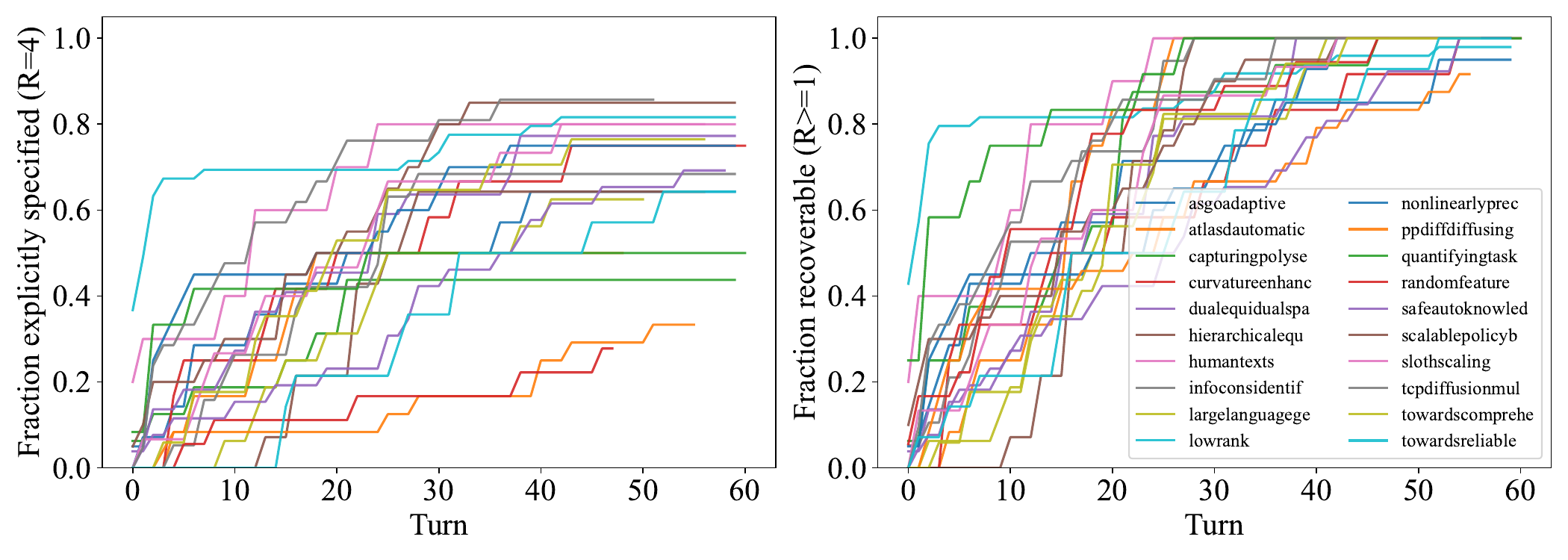}
    \caption{Cumulative exposure over turns. \textbf{Left:} fraction of components explicitly specified ($R{=}4$). \textbf{Right:} fraction recoverable ($R{\ge}1$). Each line is one paper. The gap between panels reflects the ambiguous middle where components are inferable but not yet fully specified. }
    \label{fig:exposure_drift}
\end{figure*}

\section{Benchmark}
\label{sec:benchmark}

Our benchmark is designed to evaluate specification tracking under emergent-specification coding. It targets four properties. First, the target system should emerge through interaction rather than being given upfront in a single prompt. Second, sessions should be long enough to require the agent to preserve design commitments across a sustained implementation arc rather than only across local edits. Third, evaluation should be component-level and tied to a frozen target specification derived from the source paper. Fourth, the papers should be recent and high-value so that contamination risk is reduced, though not eliminated. The benchmark therefore aims to be neither trivially unfair nor trivially explicit: most scored components must be recoverable from the visible interaction, but many should only become clear after integrating information across turns.

We construct the benchmark from 20 recent ML papers: 10 randomly sampled from ICML~2025 and 10 from NeurIPS~2025. Each paper is converted into an emergent-specification coding task together with a frozen component checklist for evaluation.

\subsection{Construction Pipeline}
\label{subsec:construction}

All benchmark assets are produced with a single frozen model snapshot (\textsc{claude-sonnet-4-5-20250929}) using fixed prompt templates and schema-constrained outputs. The model serves as a controlled annotation instrument rather than an oracle for scientific truth: the original paper remains the sole authority for all paper-grounded content. Exact prompts, snapshot identifiers, and output schemas are provided in the appendix.

\paragraph{Phase 1: specification grounding.}
For each paper, we first extract a canonical specification from the PDF. This artifact contains a detailed method description with explicit mathematical content, the datasets and metrics used in evaluation, a step-by-step implementation plan, and plausible local design alternatives for major components. The purpose of the canonical specification is not to reconstruct the authors' historical development process, but to define a single frozen implementation target against which final repositories can be scored.

We then decompose the canonical specification into atomic verifiable components. Each component records its intended functionality, exact formula when applicable, expected inputs and outputs, and module-level connections to the rest of the system. We focus on algorithmic structure and inter-module wiring. Hyperparameters are excluded unless they are part of the mathematical identity of the method. This extraction is performed once per paper and frozen before any agent run. Across the 20 papers, the benchmark contains 371 verifiable components in total.

\paragraph{Phase 2: interaction scaffolding.}
Given the canonical specification, we synthesize a latent trajectory of 10 to 20 intermediate project versions. The first version begins from a deliberately underspecified idea, such as a loose objective family or a vague architectural template, and the final version matches the canonical method exactly. Each transition introduces one meaningful refinement to the design. We then convert these refinements into natural-language coding requests under a strict agent-blindness constraint: the agent sees neither the source paper nor the hidden canonical specification. Each paper yields approximately 60 user requests.

These trajectories are synthetic rather than historical. They are not intended to reconstruct how the original authors discovered the method. Instead, they are constrained to satisfy three benchmark requirements: temporal coherence, meaningful refinement at each step, and exact convergence to the canonical specification. This lets us evaluate whether an agent can recover and preserve a distributed target design without ever seeing that design in one place.

\subsection{Scoring Protocol}
\label{subsec:scoring}

After each run, the final repository is scored component-wise against the frozen checklist using the five-level rubric in Table~\ref{tab:benchmark_rubric}. Scoring is performed by a bounded LLM judge with at most eight repository-inspection tool calls per component. The judge is given the component specification and may inspect the repository, but it does not have access to hidden benchmark artifacts beyond the frozen component description.

The distinction between \emph{equivalent} and \emph{faithful} is important. A component receives score~3 if the implementation differs in form but preserves the same functional behavior and system role. It receives score~4 only when the implementation closely matches both the intended formulation and its architectural realization. This distinction matters because \phenomenon{} is not limited to outright omission or failure; it can also appear as a structurally weaker or differently realized implementation that still partially works.

\begin{table}[t]
\centering
\small
\begin{tabular}{c l p{0.48\linewidth}}
\toprule
Score & Label & Definition \\
\midrule
0 & Absent & No code attempts this component. \\
1 & Wrong & Code attempts the component but implements logic inconsistent with the specification. \\
2 & Simplified & Code implements a degraded or approximate version of the intended component. \\
3 & Equivalent & Code differs in form but preserves the same functional behavior and system role. \\
4 & Faithful & Code closely matches the intended formulation and architectural realization. \\
\bottomrule
\end{tabular}
\caption{Component-level faithfulness rubric used to score final repositories against the frozen benchmark specification.}
\label{tab:benchmark_rubric}
\end{table}

\subsection{Benchmark Fairness Validation}
\label{subsec:validity}

An emergent-specification benchmark raises an immediate validity question: are the scored components actually recoverable from the interaction the agent observed? If not, low scores would reflect benchmark unfairness rather than a failure of specification tracking. We address this with an exposure audit conducted independently of any agent run.

For each component $c$, we annotate the recoverability level from the visible interaction and repository state alone. The recoverability level $R(c) \in \{0,1,2,3,4\}$ records whether a competent agent could infer the intended component from the visible interaction. Score~0 indicates that the component is not recoverable from the visible interaction. Score~1 indicates weak recoverability. Score~2 indicates recoverability with residual ambiguity. Score~3 indicates clear recoverability. Score~4 indicates that the component is explicitly specified.

We summarize this audit with two corpus-level statistics. The Recoverable Component Rate is
\[
\mathrm{RCR} = \frac{1}{|C|}\sum_{c}\mathbf{1}[R(c)\geq 1],
\]
the fraction of scored components that are at least weakly inferable from the visible interaction. The Explicitly Specified Rate is
\[
\mathrm{ESR} = \frac{1}{|C|}\sum_{c}\mathbf{1}[R(c)=4],
\]
the fraction stated without ambiguity. Across the 20 papers, the macro-averaged RCR is 0.994, while the macro-averaged ESR is 0.666. In total, 358 of 371 components are both committed and recoverable. Thus, nearly all scored components are available from the visible dialogue, but many must be assembled across turns rather than copied from a single explicit instruction.

Figure~\ref{fig:exposure_drift} plots the cumulative fraction of components that are explicitly specified ($R{=}4$, left) and recoverable ($R{\ge}1$, right) as a function of turn number, for each paper.
Two features are visible.
First, the curves are spread across papers: some methods are largely specified by turn~20, while others continue introducing new components past turn~40.
Second, the right panel converges faster than the left, indicating that many components enter an ambiguous window (recoverable but not yet explicit) before their final form is committed.
This temporal structure is by design and ensures that the benchmark stresses long-horizon retention rather than testing only the agent's response to the final few turns.

\subsection{Score Calibration}
\label{subsec:calibration}
\paragraph{Benchmark-asset audit.}
We sample five papers and manually inspect their canonical specifications, latent trajectories, generated request scripts, and extracted component sets against the original papers.
The audit checks four properties: canonical specifications preserve the paper's mathematical content without hallucinating unsupported commitments; design alternatives are plausible local options rather than historical claims; the latent trajectory is temporally coherent and converges exactly to the canonical specification; and the coding requests are sufficiently specific for an agent that cannot access hidden benchmark artifacts.
Based on this audit, we revise the prompt templates once and freeze the final construction pipeline.
\paragraph{Judge calibration.}
Because both benchmark construction and scoring involve LLM-based annotation, we calibrate the automated judge against human labels~\citep{zhuge2025agentjudge} on 120 component-run pairs stratified across papers, platforms, and score levels.  Two independent annotators, each blind to the judge and to each other, inspect the repository and assign scores using the same five-level rubric. Table~\ref{tab:calibration} reports agreement between each annotator and the automated judge, as well as inter-annotator agreement. Agreement is high: weighted Cohen's $\kappa$ is 0.882--0.927 between annotators and judge, inter-annotator agreement is 0.930, and 96.7\%--99.2\% of pairwise scores fall within $\pm 1$. The dominant residual error is conservative false absence: in four cases per annotator, the judge assigns score~0 despite relevant code existing in the repository, with three such cases shared across both annotators. This suggests that remaining disagreement is driven more by bounded repository search than by rubric ambiguity.

 Full details about the benchmark can be found in Appendix~\ref{app:benchmark_construction}.
 
\begin{table}[t]
\centering
\small
\begin{tabular}{l c c c}
\toprule
Metric & A1--Judge & A2--Judge & A1--A2 \\
\midrule
Weighted $\kappa$ (quadratic) & 0.927 & 0.882 & 0.930 \\
Exact-match accuracy & 91.7\% & 86.7\% & 88.3\% \\
Spearman $\rho$ & 0.938 & 0.899 & 0.932 \\
Within $\pm 1$ & 97.5\% & 96.7\% & 99.2\% \\
\midrule
Boundary disagr.\ (2--3) & 1.8\% & 4.9\% & 5.2\% \\
Boundary disagr.\ (3--4) & 3.3\% & 6.2\% & 7.9\% \\
\bottomrule
\end{tabular}
\caption{Pairwise agreement among the automated component judge and two independent annotators (each blind to the other scores) on 120 stratified component-run pairs.}
\label{tab:calibration}
\end{table}

\section{Experiments}
\label{sec:experiments}

\subsection{Setup}
\label{subsec:setup}

We evaluate whether emergent specification reduces final implementation faithfulness relative to a single-shot specification control. We use two commercial long-horizon coding systems, Claude Code and Codex. For each of the 20 benchmark papers, we run two conditions on each platform. In the emergent condition, the agent receives the multi-turn interaction script and must recover the target design progressively through dialogue. In the single-shot specification control, the agent receives the complete paper in one prompt at the start of the session. This yields 80 runs in total.

The single-shot specification control is the relevant contrast for \phenomenon{}. It keeps the platform, model family, and tooling fixed while removing the need to assemble the design across turns. The resulting gap therefore measures the cost of progressive specification disclosure under the same coding environment.

We report three metrics. Mean Component Faithfulness (MCF) measures semantic faithfulness to the committed design. Let $s_i \in \{0,1,2,3,4\}$ be the rubric score for component $i$ and let $N$ be the number of components in the paper. Then
\[
\mathrm{MCF} = \frac{1}{N}\sum_{i=1}^{N} s_i.
\]
MCF ranges from 0 to 4.

Dependency Integration Ratio (DIR)~\citep{le2025impacts} measures structural integration:
\[
\mathrm{DIR} = \frac{|E_{\mathrm{used}}|}{|E_{\mathrm{total}}|},
\]
where $E_{\mathrm{total}}$ is the set of public symbols exported by standalone modules and $E_{\mathrm{used}}$ is the subset actually reused by downstream integration code. DIR ranges from 0 to 1.

To summarize both dimensions, we define implementation faithfulness
\[
\mathrm{IF50} = \frac{1}{2}\left(\frac{\mathrm{MCF}}{4}\right) + \frac{1}{2}\mathrm{DIR}.
\]
IF50 ranges from 0 to 1 and serves as our primary endpoint metric for \phenomenon{}.

\begin{figure}
    \centering
    \includegraphics[width=\linewidth]{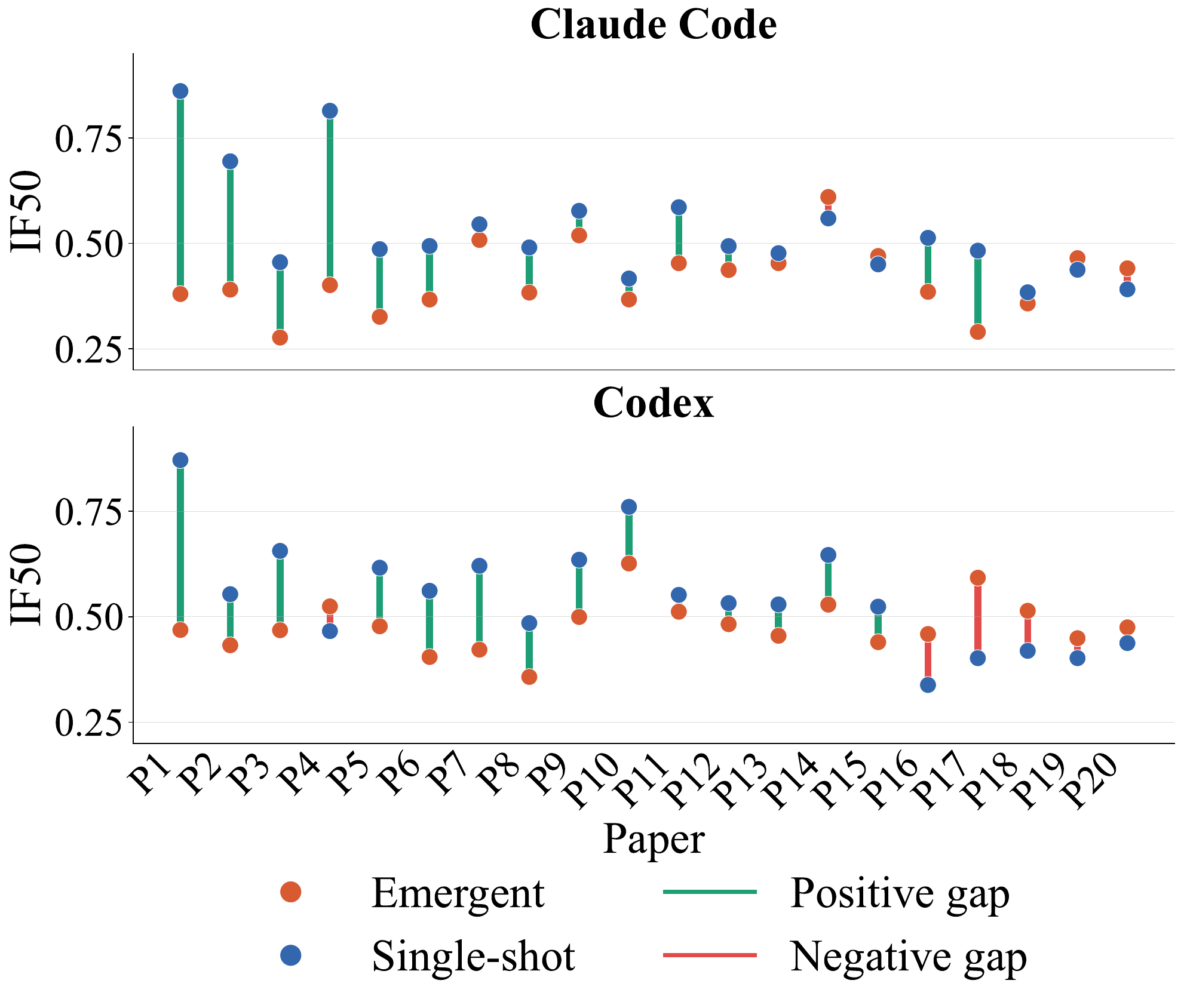}
    \caption{Per-paper IF50 under emergent
specification and the single-shot specification control. }
    \label{fig:if50_dumbbell}
\end{figure}

\subsection{Results}
\label{subsec:results}

Figure~\ref{fig:if50_dumbbell} shows per-paper IF50 under emergent specification and the single-shot specification control. On Claude Code, the single-shot control outperforms emergent specification on 16 of 20 papers, with a mean IF50 gap of $+0.116$ and Wilcoxon signed-rank $p = 0.0003$. On Codex, the same pattern holds on 14 of 20 papers, with a mean gap of $+0.071$ and $p = 0.012$. This is the primary evidence for \phenomenon{}: when the same platform receives the full design upfront rather than progressively through interaction, final implementation fidelity is higher.

The decomposition of this gap differs across platforms. On Claude Code, emergent specification lowers both semantic faithfulness and structural integration: MCF drops from 3.031 under the single-shot control to 2.718 under emergent specification, while DIR drops from 0.303 to 0.149. On Codex, the MCF gap is negligible (3.245 versus 3.242), but the DIR gap remains substantial (0.289 versus 0.148). Thus, \phenomenon{} is robust at the overall IF50 level and robust on structural integration across both platforms, while semantic faithfulness loss is concentrated on Claude Code.

This platform asymmetry is important. It shows that \phenomenon{} is not tied to a single narrow failure pattern. Progressive specification disclosure can reduce final implementation fidelity through different mixtures of semantic loss and structural non-reuse on different systems. For this reason, we treat MCF and DIR as diagnostic views of \phenomenon{}, while using IF50 as the primary summary measure.

\begin{figure}
    \centering
    \includegraphics[width=.7\linewidth]{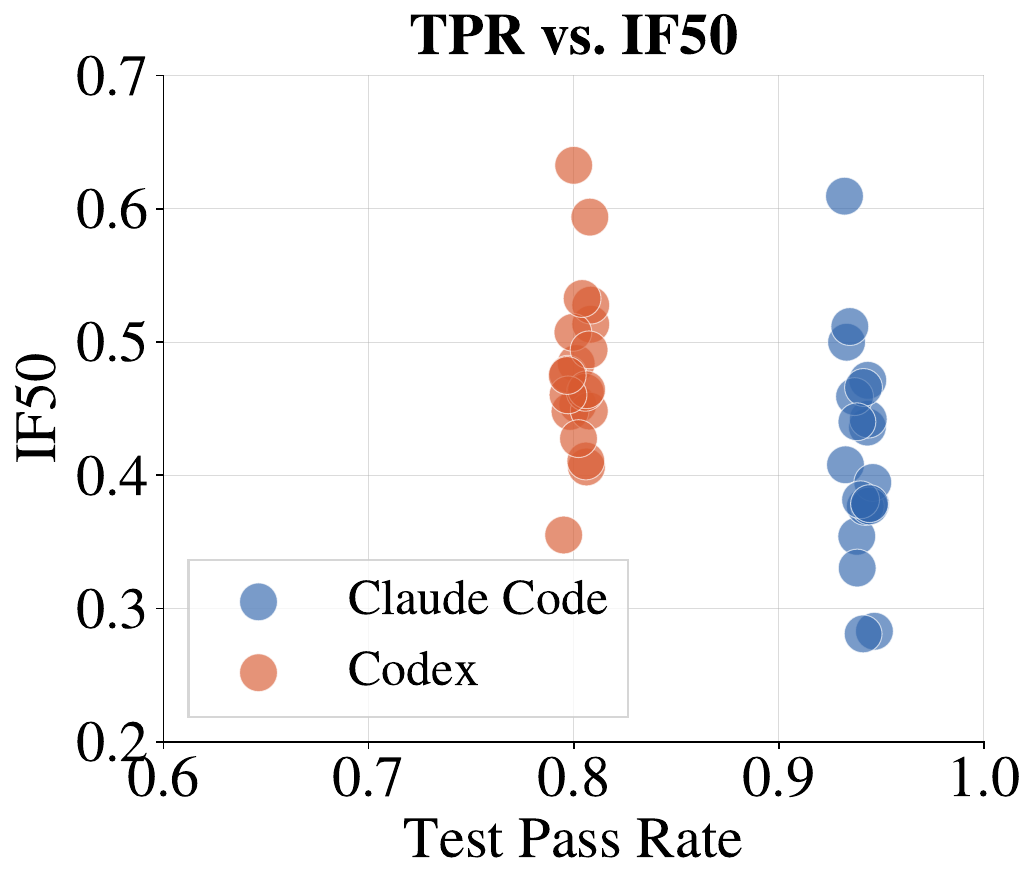}
    \caption{Test pass rate fails to detect \phenomenon{}.}
    \label{fig:tpr_dissociation}
\end{figure}

Note that executable tests do not reliably detect \phenomenon{}. Across multi-turn runs, as shown in Figure~\ref{fig:tpr_dissociation}, the correlation between test pass rate and IF50 is low, indicating that agent-authored tests may validate the implementation that was produced without verifying that it remains faithful to the intended design.

\begin{table}[t]
\centering
\small
\begin{tabular}{lccc}
\toprule
Platform / Condition & MCF & DIR & IF50 \\
\midrule
Claude Code / Emergent & 2.718 & 0.149 & 0.414 \\
Claude Code / Single-shot & 3.031 & 0.303 & 0.530 \\
Codex / Emergent & 3.242 & 0.148 & 0.479 \\
Codex / Single-shot & 3.245 & 0.289 & 0.550 \\
\bottomrule
\end{tabular}
\caption{Aggregate faithfulness under emergent specification and the single-shot specification control.}
\label{tab:main_results}
\end{table}

\paragraph{Qualitative example.}
One representative example comes from the DualEquiNet task~\citep{xu2025dualequinet}. By turn~40/97, the visible interaction clearly commits to a three-stream residual update: the scalar state receives the sum of the EU and SH deltas, coordinates receive an EU coordinate delta, and SH features receive an SH delta. In the emergent run, the agent initially implements this design as a dedicated module (\texttt{equivariant\_coordinate\_update.py}), but the final consolidated model later bypasses that module, drops the coordinate update, and replaces the summed scalar residual with separate \texttt{h\_eu}/\texttt{h\_sh} MLP paths. The final repository therefore receives a \emph{Simplified} score on this component, and the earlier module is no longer reused by the final pipeline. Under the single-shot specification control, the same platform preserves all three residual paths in a single layer module (\texttt{src/layers/dualequi\_layer.py}), yielding a \emph{Faithful} score. This example illustrates \phenomenon{} as a failure to preserve a committed component during integration rather than a failure to implement it in isolation.

\begin{table}[t]
\centering
\small
\begin{tabular}{p{0.18\linewidth} p{0.68\linewidth}}
\toprule
Case & Description \\
\midrule
Committed design & Three-stream residual update committed in the visible interaction by turn~40/97: scalar state receives summed EU+SH deltas; coordinates receive an EU delta; SH features receive an SH delta. \\\\
Emergent & Final integrated model bypasses the earlier module, drops the coordinate update, and replaces the summed scalar residual with separate \texttt{h\_eu}/\texttt{h\_sh} paths. Score: \emph{Simplified} (2/4). \\\\
Single-shot & Final model preserves all three residual paths in one layer module, including the coordinate update and summed scalar residual. Score: \emph{Faithful} (4/4). \\
\bottomrule
\end{tabular}
\caption{Representative example of \phenomenon{} on a clearly committed component. The emergent run implements the intended module earlier in the session but does not preserve it in the final integrated pipeline, whereas the single-shot specification control does.}
\label{tab:qual_case}
\end{table}
\section{Mitigating \phenomenon{} with \memory{}}
\label{sec:memory}

Section~\ref{sec:experiments} showed that emergent specification reduces final implementation faithfulness relative to the single-shot specification control, with losses visible in both semantic faithfulness and structural integration. As a case study in targeted mitigation, we introduce \memory{}, an external project-state layer designed to improve specification tracking in long-horizon coding without modifying the underlying model or the platform's native context-management mechanism.

\subsection{Method}
\label{subsec:memory_architecture}

\memory{} maintains two synchronized external views of the project. The first is a semantic state that stores durable project knowledge, including design commitments, implementation decisions, and supporting resources extracted from accepted project progress. This state is conservative: it records committed project facts rather than every proposal made during the interaction. The second is a structural state that summarizes the current repository organization, including files, public symbols, and interface relations derived directly from the codebase. The semantic state captures what the project is intended to do; the structural state captures where those commitments live in code and what compatibility constraints later edits must respect.

Before each coding turn, a forecaster combines the current request with prior user requests and the external project state to generate a task brief for the coding agent. The brief highlights relevant design commitments, identifies modules that should be preserved or revised, and surfaces repository constraints needed for compatibility-aware modification. When the remaining live context is likely insufficient for the upcoming request, \memory{} starts a fresh session and injects a rendered project brief containing the current semantic and structural state.

This design targets the two diagnostic dimensions of \phenomenon{}. The semantic state supports retention of committed algorithmic content, while the structural state and forecasting step support reuse and integration of previously implemented modules.

\subsection{Results}
\label{subsec:pg_results}

We evaluate whether \memory{} reduces \phenomenon{} by comparing emergent+\memory{} runs against the plain emergent baseline on the same benchmark and platforms. To contextualize the gains, we also report gap recovery relative to the full-specification control:
\[
\mathrm{Recovery}_F =
\frac{F(R^{\mathrm{mem}})-F(R^{\mathrm{em}})}
     {F(R^{\mathrm{ss}})-F(R^{\mathrm{em}})},
\]
when the denominator is positive.

On Claude Code, where Section~\ref{subsec:results} showed substantial \phenomenon{}, \memory{} improves MCF from 2.718 to 3.000, recovering 90\% of the full-to-emergent gap. Fully faithful components increase from 118 to 181 out of 371, while severe failures decrease from 72 to 49. IF50 improves on 15 of 20 papers. These gains indicate that external project state can substantially reduce faithfulness loss when the baseline platform exhibits it.

On Codex, semantic gains are small, consistent with the limited MCF gap observed in Section~\ref{subsec:results}. Structural gains remain meaningful: DIR improves on 15 of 19 papers with non-missing values, with mean relative improvement of 76\%. This suggests that compatibility-aware forecasting and structural state can improve integration even when semantic faithfulness is already comparatively stable.

Per-paper results, overhead statistics, and implementation details are provided in the Appendix~\ref{app:memory}.

\section{Related Work}
\label{sec:related-work}
Recent work benchmarks coding agents under interactive, project-level, or research-oriented settings. InterCode~\citep{yang2023intercode} introduced interactive coding with execution feedback. ProjectEval~\citep{projecteval}, RepoExec~\citep{le2025impacts}, and FEA-Bench~\citep{li2025feabench} evaluate project- and repository-level code generation from fixed task descriptions or provided repository context. In research-oriented settings, PaperBench~\citep{starace2025paperbench}, LMR-BENCH~\citep{lmrbench}, ResearchCodeBench~\citep{hua2025researchcodebench}, and RECODE-H~\citep{miao2025recode} evaluate paper-grounded implementation when the paper, masked repository targets, or structured task feedback are available to the agent from the start, while MLR-Bench~\citep{chen2025mlrbench} broadens the setting to open-ended ML-research workflows beyond coding alone. Closer in spirit, CodeFlowBench~\citep{wang2025codeflowbench} and SR-Eval~\citep{zhan2025sr} study multi-turn code generation under iterative decomposition or stepwise requirement refinement. Our benchmark differs in one specific way: the target design is paper-grounded but never shown upfront, and evaluation targets final faithfulness to the accumulated design rather than completion against a fixed initial specification.

A separate literature studies how agents cope with long horizons and large repositories. At the context level, prior work explores simple masking versus summarization, gist-style memory, activation-level compression, and learned context-management mechanisms such as Context Folding~\citep{sun2025contextfolding}, AgentFold~\citep{ye2025agentfold}, and MEM1~\citep{zhou2026mem1}. At the repository level, Think-Search-Patch~\citep{xiong2025thinksearchpatch}, repository-memory methods, and structure-aware retrieval methods improve code repair or repository code generation by recovering relevant context from files, commits, issues, or structural graphs. Concurrently, SWE-ContextBench~\citep{zhu2026swecontextbench} and LoCoEval~\citep{liu2026scalable} evaluate context reuse and retention in repository-oriented task sequences. Our setting is complementary: the state that must be tracked is not only latent in the repository or earlier observations, but also newly committed design knowledge introduced by the user over time.

Finally, several works argue that execution-based metrics alone are too coarse for agentic development. Agent-as-a-Judge~\citep{zhuge2025agentjudge} provides requirement-level evaluation for development tasks, LLM critics for code changes study execution-free evaluation of repository patches, and SciCoQA~\citep{baumgaertner2026scicoqa} focuses specifically on discrepancies between scientific papers and codebases. Our evaluation follows the same general motivation but targets a different object: faithfulness of the final repository to a progressively disclosed, paper-derived design.

\section{Conclusion}
\label{sec:conclusion}

We introduced a benchmark to measure \phenomenon{}, the reduction in final implementation faithfulness when a target design is progressively disclosed rather than provided upfront. Among 20 papers in the benchmark, emergent specification consistently lowers implementation fidelity relative to a single-shot specification control, especially in structural integration. A mitigation case study shows that explicitly tracking external semantic and structural project state substantially reduces \phenomenon{}. The benchmark and results highlight specification tracking as a distinct challenge for evaluating and improving long-horizon coding agents.
\clearpage
\section*{Limitations}
\label{sec:limitations}

This work has several limitations. First, the benchmark is built from 20 recent ML papers and instantiated on two commercial long-horizon coding platforms. Although this is sufficient to establish the existence of \phenomenon{} in our setting, it does not by itself characterize all research domains, all task types, or all coding agents. In particular, the benchmark focuses on method implementation with substantial mathematical structure, and the observed patterns may differ for broader software-engineering tasks or for research workflows in which the final design is less crisply defined.

Second, the benchmark interaction scripts are synthetic. They are constructed to be temporally coherent and to converge exactly to the canonical specification, but they do not reconstruct the authors' historical research process. Our exposure audit helps validate that the scored components are recoverable from the visible interaction, yet synthetic trajectories may still differ from real user behavior in pacing, ambiguity, and revision style.

Third, \phenomenon{} is defined here as an endpoint gap relative to a single-shot specification control. We do not measure temporal degradation within a session, and we do not claim that our experiments isolate a unique mechanism behind the observed loss. The diagnostic dimensions in this paper, semantic faithfulness and structural integration, are useful views of the effect, but they are not a complete causal account.

Fourth, both benchmark construction and component scoring rely in part on LLM-based annotation. We mitigate this with frozen prompts, a bounded judge, an exposure audit, and calibration against blinded human annotations, but residual annotation error remains possible. In addition, the single-shot specification control should be interpreted as a matched comparison condition, not as a universal upper bound on achievable performance.

Finally, \memory{} is presented as a case study in targeted mitigation rather than as a definitive solution. Its design choices are motivated by the benchmark's diagnostic dimensions, but the current experiments do not establish which components of the system are necessary, whether the approach generalizes across more platforms, or how it compares with a broader range of memory and context-management baselines.
\clearpage
\bibliography{custom}
\clearpage
\appendix
 
\section{Benchmark Construction Details}
\label{app:benchmark_construction}

 \subsection{Sampling Frame and Paper List}
\label{app:paper_list}

We construct the benchmark from 20 recent ML papers, randomly sampling 10 from ICML~2025 and 10 from NeurIPS~2025. We do not further curate the sample for benchmark difficulty. This choice intentionally preserves natural variation across papers in mathematical complexity, component count, interaction length, and measured \phenomenon{}. The benchmark therefore reflects the heterogeneity of recent conference methods rather than a hand-selected set of particularly clean or difficult papers.

Table~\ref{tab:paper_list} lists the sampled papers together with their component counts and interaction lengths. Below we provide a brief description of each paper to clarify the domain and methodological character of the corresponding benchmark task.

\begin{table*}[t]
\centering
\small
\begin{tabular}{l l p{0.54\linewidth} c c}
\toprule
ID & Venue & Paper Title & \# Comp. & \# Req. \\
\midrule
P1  & NeurIPS~2025 & ASGO: Adaptive Structured Gradient Optimization & 14 & 60 \\
P2  & ICML~2025    & AtlasD: Automatic Local Symmetry Discovery & 12 & 49 \\
P3  & NeurIPS~2025 & Capturing Polysemanticity with PRISM: A Multi-Concept Feature Description Framework & 16 & 60 \\
P4  & ICML~2025    & Curvature Enhanced Data Augmentation for Regression & 12 & 61 \\
P5  & NeurIPS~2025 & DualEqui: A Dual-Space Hierarchical Equivariant Network for Large Biomolecules & 22 & 60 \\
P6  & ICML~2025    & Hierarchical Equivariant Policy via Frame Transfer & 14 & 57 \\
P7  & NeurIPS~2025 & Human Texts Are Outliers: Detecting LLM-generated Texts via Out-of-distribution Detection & 10 & 60 \\
P8  & ICML~2025    & InfoCons: Identifying Interpretable Critical Concepts in Point Clouds via Information Theory & 19 & 60 \\
P9  & ICML~2025    & Large Language-Geometry Model: When LLM meets Equivariance & 16 & 51 \\
P10 & NeurIPS~2025 & Low Rank Gradients and Where to Find Them & 49 & 60 \\
P11 & ICML~2025    & Nonlinearly Preconditioned Gradient Methods under Generalized Smoothness & 20 & 60 \\
P12 & ICML~2025    & PPDiff: Diffusing in Hybrid Sequence-Structure Space for Protein-Protein Complex Design & 24 & 56 \\
P13 & NeurIPS~2025 & Quantifying Task-relevant Similarities in Representations Using Decision Variable Correlations & 12 & 61 \\
P14 & ICML~2025    & Random Feature Representation Boosting & 18 & 48 \\
P15 & ICML~2025    & SafeAuto: Knowledge-Enhanced Safe Autonomous Driving with Multimodal Foundation Models & 26 & 59 \\
P16 & NeurIPS~2025 & Scalable Policy-Based RL Algorithms for POMDPs & 20 & 60 \\
P17 & NeurIPS~2025 & Sloth: Scaling Laws for LLM Skills to Predict Multi-Benchmark Performance across Families & 15 & 57 \\
P18 & ICML~2025    & TCP-Diffusion: A Multi-modal Diffusion Model for Global Tropical Cyclone Precipitation Forecasting with Change Awareness & 21 & 52 \\
P19 & NeurIPS~2025 & Towards Comprehensive Scene Understanding: Integrating First and Third-Person Views for LVLMs & 17 & 57 \\
P20 & NeurIPS~2025 & Towards Reliable Code-as-Policies: A Neuro-Symbolic Framework for Embodied Task Planning & 14 & 60 \\
\midrule
\multicolumn{3}{l}{Mean / Total} & 18.6 / 371 & 57.4 / 1148 \\
\bottomrule
\end{tabular}
\caption{Benchmark paper list. The sample comprises 10 ICML~2025 and 10 NeurIPS~2025 papers selected at random. Component counts range from 10 to 49 (mean 18.6); interaction scripts range from 48 to 61 requests (mean 57.4).}
\label{tab:paper_list}
\end{table*}

\paragraph{P1: ASGO~\cite{an2025asgo}.}
Proposes Adaptive Structured Gradient Optimization, which exploits the low-rank and block-diagonal structure of gradient matrices to reduce the memory and computational cost of second-order optimization methods. The implementation involves structured preconditioner construction and adaptive rank selection.

\paragraph{P2: AtlasD~\cite{bhat2025atlasd}.}
Introduces an algorithm for automatically discovering local symmetries in data by learning group-equivariant mappings without requiring prior specification of the symmetry group. The benchmark components center on symmetry detection modules and equivariant network layers.

\paragraph{P3: PRISM~\cite{kopf2025capturing}.}
Presents a multi-concept feature description framework for neural network interpretability that captures polysemantic neurons by decomposing activations into multiple semantically distinct concepts rather than assigning a single label per unit.

\paragraph{P4: Curvature Enhanced Data Augmentation~\cite{sirot2025curvature}.}
Develops a data augmentation strategy for regression tasks that leverages curvature information from the input manifold to generate synthetic training samples in regions where the target function varies most rapidly.

\paragraph{P5: DualEquiNet~\cite{xu2025dualequinet}.}
Proposes a dual-space hierarchical equivariant network for modeling large biomolecular structures by operating simultaneously in Cartesian and internal coordinate spaces. The architecture enforces SE(3) equivariance through hierarchical message passing across spatial scales.

\paragraph{P6: Hierarchical Equivariant Policy~\cite{zhao2025hierarchical}.}
Introduces a robot manipulation policy that achieves SE(3) equivariance through frame transfer across a hierarchy of reference frames, enabling sample-efficient learning of spatial reasoning for object rearrangement and assembly tasks.

\paragraph{P7: Human Texts Are Outliers~\cite{zeng2025human}.}
Formulates LLM-generated text detection as an out-of-distribution detection problem, treating human-written text as the in-distribution class and leveraging distributional divergence measures to identify machine-generated content without requiring access to the generating model.

\paragraph{P8: InfoCons~\cite{li2025infocons}.}
Proposes an information-theoretic framework for identifying interpretable critical concepts in point cloud data by selecting subsets of geometric primitives that maximize mutual information with downstream task labels.

\paragraph{P9: Large Language-Geometry Model~\cite{li2025largelanguagegeometrymodelllm}.}
Bridges large language models and geometric reasoning by incorporating equivariant representations into the language model architecture, enabling joint processing of natural language instructions and three-dimensional molecular or spatial inputs.

\paragraph{P10: Low Rank Gradients~\cite{sonthalia2025lowrankgradients}.}
Provides theoretical and empirical analysis of when and why gradient matrices in deep networks exhibit low-rank structure during training, and develops practical criteria for predicting gradient rank as a function of architecture and training dynamics. This paper yields the largest number of verifiable components (49) in the benchmark.

\paragraph{P11: Nonlinearly Preconditioned Gradient Methods~\cite{oikonomidis2025nonlinearlypreconditionedgradientmethods}.}
Extends preconditioned gradient descent to settings with generalized smoothness conditions, providing convergence guarantees for nonlinear preconditioning strategies that go beyond the standard Lipschitz gradient assumption.

\paragraph{P12: PPDiff~\cite{song2025ppdiffdiffusinghybridsequencestructure}.}
Introduces a diffusion model for protein-protein complex design that operates in a hybrid sequence-structure space, jointly generating amino acid sequences and three-dimensional backbone conformations for protein-protein interfaces.

\paragraph{P13: Quantifying Task-Relevant Similarities~\cite{qian2025quantifying}.}
Proposes decision variable correlations as a metric for comparing neural representations in terms of their task-relevant content, offering a behaviorally grounded alternative to representational similarity analysis.

\paragraph{P14: Random Feature Representation Boosting~\cite{zozoulenko2025randomfeaturerepresentationboosting}.}
Develops a boosting framework over random feature representations, iteratively constructing an ensemble of random feature models whose combined representation improves approximation quality for kernel methods.

\paragraph{P15: SafeAuto~\cite{zhang2025safeautoknowledgeenhancedsafeautonomous}.}
Integrates structured driving safety knowledge into multimodal foundation models for autonomous driving, using knowledge graphs and retrieval mechanisms to enforce safety constraints during planning and decision making.

\paragraph{P16: Scalable Policy-Based RL for POMDPs~\cite{anjarlekar2025scalablepolicybasedrlalgorithms}.}
Derives scalable policy gradient algorithms for partially observable Markov decision processes by exploiting structural properties of belief-space optimization to reduce the variance and computational cost of gradient estimation.

\paragraph{P17: Sloth~\cite{polo2025slothscalinglawsllm}.}
Proposes scaling laws that decompose LLM performance into skill-specific components, enabling prediction of multi-benchmark performance across model families from a small number of representative evaluations.

\paragraph{P18: TCP-Diffusion~\cite{huang2025tcpdiffusionmultimodaldiffusionmodel}.}
Presents a multimodal diffusion model for forecasting tropical cyclone precipitation, incorporating change-awareness mechanisms that condition the generative process on temporal evolution patterns in satellite and reanalysis data.

\paragraph{P19: Towards Comprehensive Scene Understanding~\cite{lee2025comprehensivesceneunderstandingintegrating}.}
Addresses the integration of first-person and third-person visual perspectives within large vision-language models, proposing architectural modifications that enable joint reasoning across viewpoints for scene understanding tasks.

\paragraph{P20: Towards Reliable Code-as-Policies~\cite{ahn2025reliablecodeaspoliciesneurosymbolicframework}.}
Introduces a neuro-symbolic framework for embodied task planning that improves the reliability of code-as-policies approaches through formal verification of generated programs and symbolic constraint satisfaction during plan execution.

\subsection{Frozen Prompt Templates}
\label{app:benchmark_prompts}

All benchmark assets are produced using a single frozen
model snapshot with fixed prompt templates and
schema-constrained outputs. The source paper serves as
the sole authority for paper-grounded content. Below we
reproduce the exact prompt templates used in the frozen
pipeline. Placeholder variables are typeset in
\texttt{<angle brackets>}.

\medskip
\noindent\textbf{Prompt~A: Canonical Specification
Extraction.}

\begin{lstlisting}
Extract the following fields for this paper
(paper_id and title are already known; generate
only the following):

Required output schema (return ONLY valid JSON):
{
  "idea": "Summary of the method (at least 600
    words). Include key mathematical components
    such as loss functions, main equations, and
    core signals. Be specific about the
    mathematical formulation. The summary should
    be sufficiently detailed for others to
    reproduce the paper.",
  "evaluation": {
    "datasets": ["list of datasets/benchmarks"],
    "metrics": ["list of evaluation metrics"],
    "description": "How each metric is calculated"
  },
  "Implementation": "Assume you were the first
    author of this paper, starting from scratch.
    Describe step-by-step how to implement the
    idea and complete the evaluation (ignore
    baselines). Provide a concrete and detailed
    implementation document that others can
    reproduce. This is the reverse engineering
    of your research journey; do not rely on
    existing code or repositories.",
  "design_alternatives": [
    {
      "design_id": "unique identifier (e.g.,
        loss_function, architecture;
        hyperparameters are not a design)",
      "description": "What design choice this
        represents, e.g., for which component,
        in which step",
      "example_options": [
        "option1", "option2", "option3"
      ]
    }
  ]
}

Guidelines:
- Output ONLY the JSON, no additional text
\end{lstlisting}

\medskip
\noindent\textbf{Prompt~B: Atomic Component
Extraction.}

\begin{lstlisting}
You are a specification decomposer. Given a
detailed technical specification of a research
paper, decompose it into components and
identify how components are connected. For
example, how data is prepared, how the model
architecture is defined, how the loss is
backpropagated, how the output of component A
connects to component B, which subset is used
for evaluation, and how the metrics are computed.

Based on these, identify atomic verifiable claims
that can be independently checked against an
implementation. Each claim should be a single,
concrete, testable statement about what the code
MUST do.

Focus on wiring rather than concrete numbers
that are configurable.

For each claim, provide:
- description: what this component does
    (1-2 sentences)
- formula: the exact mathematical formula if
    applicable, or "N/A" if purely structural
- inputs: what data/tensors this component
    receives, their source, and their meaning
- outputs: what it produces, and what the
    outputs represent or measure
- connections: which modules this component
    should connect to, and the relationships
    between them

Output a JSON object with a single key
"components" containing an array of claims.
Example:
{"components": [
  {"description": "EU space message computation
      using invariant features and pairwise
      distances",
    "formula": "m_EU,ij = phi_EU(
      [h_i, h_j, ||x_ij||])",
    "inputs": "scalar features h_i, h_j;
      pairwise distance ||x_ij||",
    "outputs": "message vector m_EU,ij",
    "connection": "inside DualEquiLayer; output
      is sent to ..."}
]}

CRITICAL: Output the JSON object directly.
Do NOT use any tools. Do NOT write any files.
Return the JSON only.

Here is the specification:
<canonical_idea>
\end{lstlisting}

\medskip
\noindent\textbf{Prompt~C: Latent Trajectory
Generation.}

\begin{lstlisting}
Assume you were the first author of the paper
"<paper_title>" published at <venue>.

Reverse-engineer your own research journey: the
iterative process you went through before
arriving at the final published method.

Generate a sequence of project versions
(version 0 through version N) that authentically
captures this evolution.

Guidelines:
1. Start vague (version 0): High-level
   intuition, missing mathematical details,
   using generic placeholders such as "some
   contrastive loss" or "a deep network."

2. Progressive refinement: Explore at least one
   of the following actions in each subsequent
   version:
   - Brainstorm designs and ideas to achieve
     the high-level intuition of the whole
     paper or a concrete module.
   - Search for baselines and existing works.
   - Check one dependent library or benchmark
     to understand how to connect to your
     codebase.
   - Add a concrete functional module,
     detailed enough for reproduction.
   - Replace a functional module with a new
     module of better design.

3. Temporal consistency: The refinement
   process should reflect the reversed
   research journey; each refinement should
   be meaningful.

4. End at the canonical specification: The
   final version should match exactly the
   canonical specification below.

5. Be realistic: Reflect how research
   actually happens, starting with small
   steps to verify each hypothesis,
   optimizing design, examining failures,
   trying different options, and eventually
   converging.

Generate 10 to 20 versions total.

Canonical specification (your final
destination):
<canonical_spec>

Output format:
[
  {
    "version": 1,
    "Action": "Implement the core framework
      with xx model" (implementation
      description only, no code)
  },
  {
    "version": 2,
    "Action": "I noticed xx and am considering
      whether using xx would be better.
      Examine the details of xx to check
      applicability."
  },
  ...
]

Generate the evolution sequence as a JSON
array. Output ONLY the JSON array, no
additional text.
\end{lstlisting}

\medskip
\noindent\textbf{Prompt~D: Coding Request
Generation.}

\begin{lstlisting}
Assume you were the first author of
"<paper_title>". We are simulating how you
could develop and implement the paper if you
could work with a coding agent.

Your current design and project status:
<current_project_state>

You now want to make the following change:
<target_version>

Generate a sequence of requests to the coding
agent. These requests should:

1. Be clear and specific: The agent cannot see
   any of the documents above. Be clear and
   detailed about what to search, check,
   compare, or implement in each request, but
   do not write any code.

2. Show realistic task chunking: Do not
   attempt to accomplish everything at once.
   Break work into logical pieces. Each
   request can be a functional module
   involving multiple files.

3. Build sequentially: The agent will
   automatically use version control and unit
   tests to ensure correct implementation, or
   use tools to complete search and
   brainstorming tasks. Design the next
   request assuming the preceding request was
   successful, but independent of the agent
   response.

Since you were still developing the paper
step by step, you did not know whether the
approach would succeed, let alone what it
would eventually be called. Use phrases such
as "my idea" or "my hypothesis."

Generate 1 to 3 requests depending on
complexity. Output as a JSON array:
[
  {"request": "..."},
  {"request": "..."}
]

Output ONLY the JSON array, no additional
text.
\end{lstlisting}

\medskip
\noindent\textbf{Prompt~E: Bounded Repository
Judge.}

\begin{lstlisting}
You are an implementation faithfulness judge.
Determine whether a specific component is
correctly implemented in the codebase located
in your current working directory.

Specification claim to verify:

COMPONENT: <component_description>
FORMULA: <component_formula>
EXPECTED INPUTS: <expected_inputs>
EXPECTED OUTPUTS: <expected_outputs>
CONNECTED MODULE: <connected_module>

Instructions:
1. Search the codebase for code implementing
   this component (use grep, glob, read tools).
2. Compare the found code against the formula
   and description.
3. Assign a score:
   0 = ABSENT: no code attempts this component
   1 = WRONG: code attempts it but the logic
       is incorrect
   2 = SIMPLIFIED: degraded or approximate
       version (e.g., mean pooling instead of
       attention-weighted, missing formula
       terms)
   3 = EQUIVALENT: not identical but
       functionally equivalent
   4 = FAITHFUL: matches the specification

IMPORTANT: You have a LIMITED turn budget.
Perform at most 8 tool calls, then you MUST
output your JSON verdict. If you have not found
the code by then, score 0 (ABSENT). Do NOT
continue searching beyond this limit.

CRITICAL: Your FINAL message MUST end with
EXACTLY this JSON format on its own line,
with nothing after it:
{"score": <0-4>,
 "evidence": "<file:line or 'not found'>",
 "deviation": "<description or 'none'>"}
\end{lstlisting}

\subsection{Benchmark-Asset Audit}
 
We manually audited benchmark assets for five sampled papers before freezing the construction pipeline. The audit verified that the canonical specification preserved the paper's mathematical content without unsupported commitments, that component decompositions were verifiable and non-redundant, that latent trajectories were temporally coherent and converged exactly to the canonical specification, and that user-facing requests were sufficiently specific for an agent that did not see the hidden benchmark artifacts. On the basis of this audit, we revised the prompt templates once and froze the final pipeline.

\section{\memory{} Details}
\label{app:memory}

\begin{figure*}
    \centering
    \includegraphics[width=0.9\linewidth]{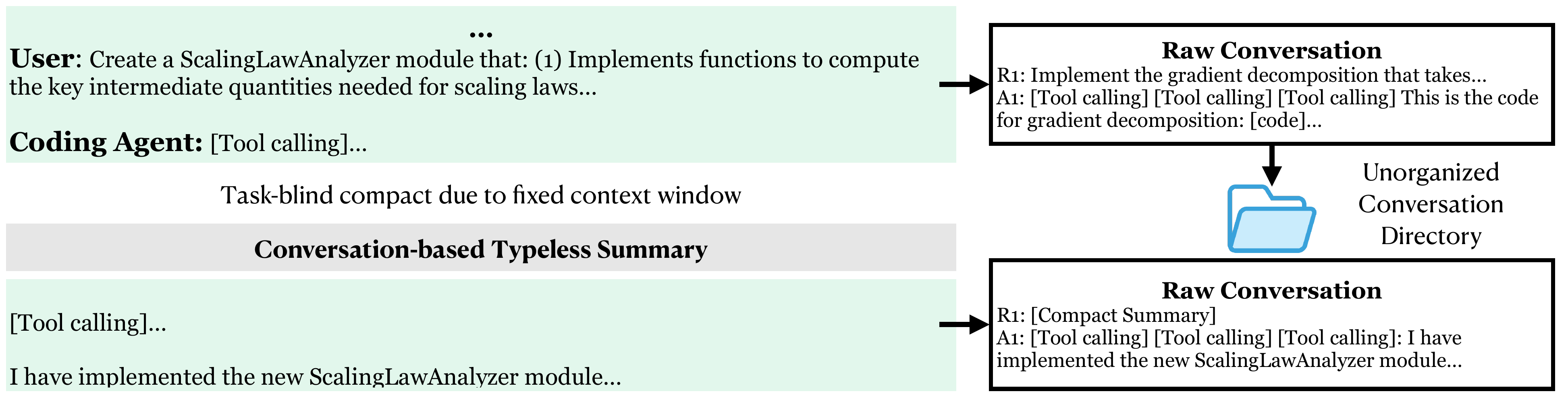}
    \caption{Default long-horizon coding workflow. User requests, tool outputs, and code edits accumulate in a single live conversation, with older context periodically condensed into a compact summary. This representation preserves recent interaction state but provides only a weak project-level view of earlier design commitments and repository structure, making specification tracking under emergent specification difficult.}
    \label{fig:default}
\end{figure*}

\begin{table*}[t]
\centering
\caption{Per-paper ProjectGuard effectiveness across 20 benchmark papers.
  MCF Recovery and IF50 Recovery are expressed as percentages of the
  specification bottleneck gap (full-paper minus emergent) closed by
  ProjectGuard.  DIR\,$\Delta$ is the relative increase in dependency
  integration ratio.  Memory is the external state size maintained by
  ProjectGuard.  $\Delta$Compaction is the change in compaction events
  per session.  Positive results (recovery or integration gain) are
  \good{highlighted}.
  CC\,=\,Claude Code;
  Papers are indexed P1--P20; --- indicates a zero denominator (no specification gap or no baseline integration).}
\label{tab:per_paper_pg}
\resizebox{\linewidth}{!}{
\begin{tabular}{@{}ll rrrrrrrrrrrrrrrrrrrr rr}
\toprule
 &
 & \rotatebox{70}{P1}
 & \rotatebox{70}{P2}
 & \rotatebox{70}{P3}
 & \rotatebox{70}{P4}
 & \rotatebox{70}{P5}
 & \rotatebox{70}{P6}
 & \rotatebox{70}{P7}
 & \rotatebox{70}{P8}
 & \rotatebox{70}{P9}
 & \rotatebox{70}{P10}
 & \rotatebox{70}{P11}
 & \rotatebox{70}{P12}
 & \rotatebox{70}{P13}
 & \rotatebox{70}{P14}
 & \rotatebox{70}{P15}
 & \rotatebox{70}{P16}
 & \rotatebox{70}{P17}
 & \rotatebox{70}{P18}
 & \rotatebox{70}{P19}
 & \rotatebox{70}{P20}
 & \rotatebox{70}{Mean} \\
\midrule
\multirow{2}{*}{MCF Rec.\ (\%)}
 & CC    & \good{+22} & $-$0 & $-$100 & \good{+150} & \good{+70} & \good{+167} & \good{+250} & \good{+100} & \good{+33} & \good{+94} & \good{+5} & \good{+36} & $-$200 & $-$75 & +0 & $-$0 & \good{+111} & +0 & \good{+200} & \good{+150} & \good{+51} \\
 & Codex & $-$600 & --- & $-$20 & \good{+33} & $-$56 & --- & \good{+29} & $-$100 & \good{+75} & $-$24 & $-$8 & \good{+32} & $-$45 & $-$33 & \good{+11} & \good{+67} & \good{+80} & \good{+57} & $-$20 & \good{+200} & $-$18  \\
\midrule
\multirow{2}{*}{DIR $\Delta$ (\%)}
 & CC    & $-$47 & $-$10 & \good{+48} & \good{+1125} & \good{+120} & \good{+955} & \good{+24} & $-$58 & \good{+13} & \good{+1} & \good{+7} & \good{+24} & $-$22 & \good{+19} & $-$8 & \good{+7} & \good{+83} & \good{+12} & $-$54 & --- & \good{+118} \\
 & Codex & \good{+15} & \good{+67} & \good{+49} & \good{+54} & \good{+55} & $-$71 & $-$52 & \good{+61} & --- & \good{+269} & $-$23 & \good{+122} & \good{+58} & \good{+171} & \good{+21} & \good{+240} & \good{+292} & \good{+19} & $-$9 & \good{+99} & \good{+76} \\
\midrule
\multirow{2}{*}{IF50 Rec.\ (\%)}
 & CC    & \good{+118} & $-$13 & \good{+5} & \good{+482} & \good{+27} & \good{+92} & \good{+142} & $-$11 & \good{+80} & \good{+104} & $-$8 & \good{+73} & \good{+355} & $-$6 & \good{+19} & \good{+13} & \good{+41} & $-$42 & \good{+15} & \good{+119} & \good{+80}  \\
 & Codex & $-$138 & $-$109 & \good{+14} & \good{+60} & $-$55 & \good{+66} & $-$71 & \good{+102} & \good{+82} & \good{+92} & \good{+5} & \good{+77} & \good{+19} & \good{+29} & $-$39 & \good{+141} & \good{+86} & $-$9 & $-$9 & \good{+68} & \good{+21} \\
\midrule
\multirow{2}{*}{Memory (KB)}
 & CC    & 395 & 171 & 297 & 196 & 631 & 233 & 592 & 398 & 193 & 1103 & 312 & 244 & 493 & 160 & 242 & 284 & 485 & 211 & 821 & 340 & 390 \\
 & Codex & 426 & 106 & 685 & 98 & 757 & 175 & 160 & 143 & 111 & 661 & 243 & 118 & 171 & 99 & 133 & 248 & 458 & 116 & 474 & 333 & 286 \\
\midrule
\multirow{2}{*}{$\Delta$Compaction}
 & CC    & $-$4 & +4 & $-$1 & +6 & $-$12 & $-$1 & $-$4 & $-$2 & +4 & $-$22 & +1 & +7 & +9 & $-$2 & +3 & +1 & $-$4 & +2 & $-$1 & $-$2 & $-$1 \\
 & Codex & +17 & +29 & +35 & +22 & +9 & +30 & +16 & +24 & +22 & +29 & +31 & +25 & +35 & +14 & +22 & +21 & +15 & +28 & +14 & +13 & +23 \\
\bottomrule
\end{tabular}
}
\end{table*}

\begin{figure*}
    \centering
    \includegraphics[width=0.9\linewidth]{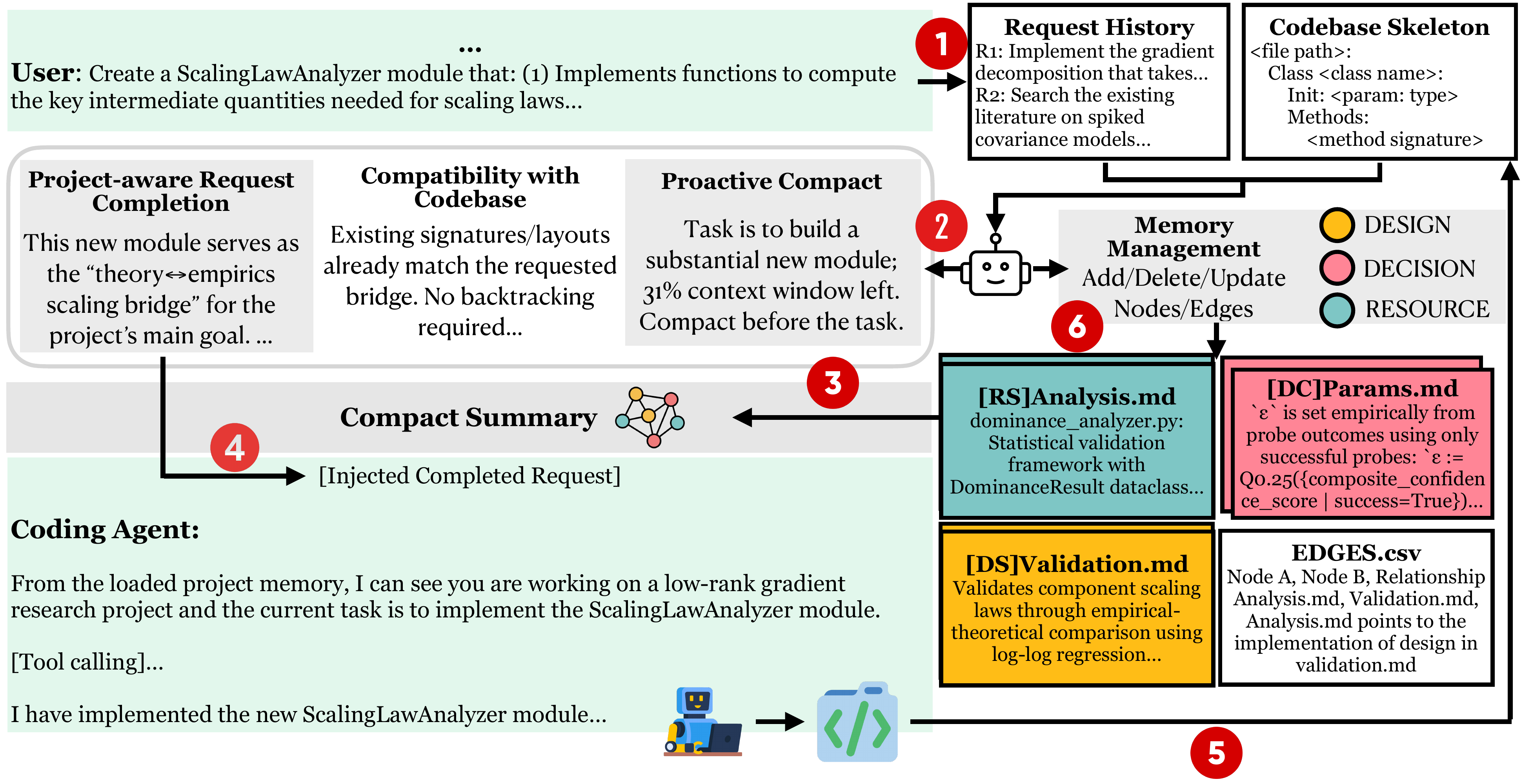}
    \caption{\memory{} overview. Before each coding turn, the system combines the request history with the current repository skeleton to maintain two external views of the project: a semantic state of committed but revisable project knowledge, and a structural state of files, symbols, and interface relations. A forecaster then prepares a project-aware brief for the coding agent, highlighting relevant design commitments, compatibility constraints, and candidate modules for revision or reuse. When the remaining live context is unlikely to support the upcoming request, \memory{} can trigger a proactive restart and inject the rendered project state into a fresh session. After the turn completes, both semantic and structural state are updated from the new interaction and repository changes.}
    \label{fig:memory}
\end{figure*}

Section~\ref{sec:problem_setting} defined \phenomenon{} as the loss in final implementation faithfulness under emergent specification relative to a single-shot specification control, and decomposed that loss into two diagnostic dimensions: semantic faithfulness and structural integration. We introduce \memory{}, an external project-state system designed to support both dimensions in long-horizon coding. Its role is not to replace the underlying coding agent, but to maintain project information outside the live context window and reintroduce that information when the current session is likely to be unreliable. The central idea is to preserve two coupled views of the project over time: a \emph{semantic state} that records revisable project knowledge, and a \emph{structural state} that records the evolving code organization needed for compatibility-aware implementation. Figures~\ref{fig:default} and~\ref{fig:memory} contrast a standard long-horizon coding workflow with our external project-state layer, illustrating how \memory{} adds explicit semantic and structural project views to support specification tracking under emergent specification.

\subsection{Semantic State}
\label{subsec:memory_semantic}

\textbf{Semantic state must remain revisable.}
In long-horizon research coding, the intended method is not specified once in a canonical prompt. It emerges through interaction and is repeatedly refined as the researcher clarifies goals, accepts some alternatives, rejects others, and revises earlier choices in light of new evidence. Supporting semantic faithfulness in this setting therefore requires more than a flat summary of prior dialogue. The system must distinguish high-level designs from lower-level implementation decisions, preserve the resources that motivate those choices, and update earlier records when the intended design changes.

Generic knowledge graphs are conceptually close to what we need, but maintaining graph state directly during many small iterative updates is cumbersome in practice. Our design therefore uses a lightweight filesystem-native graph representation that supports local updates, transparent diffs, and straightforward prompt rendering.

\textbf{Project Semantic State.}
\memory{} represents semantic project state using three record types: \emph{design}, \emph{decision}, and \emph{resource}. Design records capture higher-level method structure, architectural constraints, and other long-lived commitments. Decision records capture more specific implementation choices that operationalize a design, including interface conventions, training choices, and other details that may later need revision. Resource records capture evidence or artifacts that motivate designs and decisions, such as empirical observations and deferred questions. Each record is stored as an isolated markdown file, while relations among records are stored separately in a lightweight edge table. This representation preserves hierarchy without forcing all state into a single monolithic document.

\textbf{Semantic State Updates.}
After each coding turn, a conversation-side helper analyzes the recent interaction and extracts committed additions or revisions to semantic state. Importantly, the system is conservative about what becomes canonical. It does not treat every assistant proposal as truth. Instead, it preferentially records commitments grounded in the user's requests and in accepted project progress. A merger then updates the persistent graph, preferring edits to existing records over uncontrolled growth in the number of nodes. This is essential because long-horizon coding does not merely accumulate new facts; it frequently revises earlier ones.

\subsection{Structural State}
\label{subsec:memory_structural}

Supporting semantic faithfulness alone is not sufficient for faithful implementation. In our setting, middle turns often produce standalone modules that are individually reasonable, but later integration requires the agent to know which files already implement relevant functionality, which interfaces should be preserved, and where a design change should be propagated. Without an explicit structural view of the repository, the agent may solve each request locally, bypass earlier code, or introduce incompatible reimplementations rather than revising existing modules.

\textbf{Project Structural State.}
To support structural integration, \memory{} maintains a second external view of the project: a structural state that summarizes the repository skeleton relevant for future edits. This state includes the files that currently exist, their class and function signatures, and other interface facts that shape how modules can be reused or revised. Unlike the semantic state, which captures what the project is supposed to do, the structural state captures how the project is currently organized as code. The two are complementary: the semantic state says what should hold; the structural state says where those commitments currently live in the repository and what dependencies future edits must respect.

\textbf{Structural Updates.}
Rather than inferring structure from free-form summaries, \memory{} derives it directly from the codebase. After each coding turn, a structure-side helper checks the git history to identify which files were added, modified, or deleted, and then refreshes the project skeleton for the affected parts of the repository. This yields an up-to-date structural view grounded in actual code changes rather than in the agent's textual description of those changes.

\textbf{Compatibility-aware Forecasting.}
Before each request is executed, a forecaster reconstructs the request in project context using the history of prior user requests together with the current semantic and structural states. Its role is to recover the likely purpose of the current task, identify which earlier modules are relevant, and determine whether compatibility-preserving revision is preferable to local reimplementation. The resulting brief is passed to the coding agent so that it approaches the current task with a project-level view rather than treating it as an isolated edit.

\textbf{Proactive Restart.}
Large integration edits often require touching many existing files at once. When the forecaster estimates that the remaining live context is unlikely to support the upcoming request, it can trigger a proactive restart before the coding turn begins. In this case, \memory{} is rendered into a structured project brief and injected into a fresh coding session so that the new session begins with both semantic and structural project state available.

 \subsection{Results}
\label{subsec:pg_results}

Table~\ref{tab:per_paper_pg} reports per-paper ProjectGuard effectiveness across all 20 benchmark papers.
Recovery percentages express the fraction of the specification bottleneck gap (full-paper minus emergent) closed by ProjectGuard.

\paragraph{Claude Code.}
ProjectGuard recovers 90\% of the MCF gap, raising MCF from 2.718 (emergent) to 3.000, compared with 3.031 under the single-shot control.
The improvement is broad: 13 of 20 papers show positive MCF recovery (W/T/L = 13/4/3, median $+35$\%).
Fully faithful components (score~4) increase from 118 to 181 out of 371, a gain of 53\%, while severe failures (score~$\leq 1$) drop from 72 to 49, a reduction of 32\%.
IF50 recovery is positive on 15 of 20 papers (median $+34$\%), confirming gains in both semantic faithfulness and structural integration.

\paragraph{Codex.}
MCF improvement is negligible ($+0.05$, W/L = 9/9), consistent with the absence of a specification bottleneck gap on this platform (Section~\ref{subsec:results}).
Where no gap exists, there is nothing for ProjectGuard to recover, confirming that the MCF gains on Claude Code stem from compensating for compaction losses rather than from general coding assistance.
Structural integration tells a different story: DIR improves on 15 of 19 papers with non-missing values (mean $+76$\%, median $+55$\%), exceeding Claude Code's DIR performance (13/0/6, median $+12$\%).
The compatibility-aware forecasting and structural state components thus provide integration benefits largely independent of compaction pressure.

\paragraph{Overhead and compaction dynamics.}
ProjectGuard maintains a mean external state of 390\,KB on Claude Code and 286\,KB on Codex.
On Claude Code, compaction events per session change by a median of $-1$, indicating negligible additional context pressure.
On Codex, compaction events increase substantially (mean $+23$), but this reflects the proactive restart mechanism rather than degraded context efficiency: each restart initiates a fresh session that independently reaches Codex's compaction threshold, so the aggregate count accumulates across multiple shorter sessions rather than indicating greater pressure within any single session.
The IF50 results confirm that this increased compaction count does not erode faithfulness.

\end{document}